%Paper: hep-th/9506163
%From: dhoker@physics.ucla.edu (D'Hoker)
%Date: Sat, 24 Jun 1995 17:20:13 +0800

%
%
\magnification=\magstep1
\baselineskip=16pt
\overfullrule=0pt

\def\G{{\cal G}}
\def\H{{\cal H}}
\def\M{{\cal M}}
\def\half{{1 \over 2}}

\def\tr{{\rm tr}}
\rightline{UCLA/95/TEP/13}
\bigskip
\bigskip
\centerline{{\bf GENERAL 2+1 DIMENSIONAL EFFECTIVE ACTIONS AND}}
\medskip
\centerline{{\bf SOLITON SPIN FRACTIONALIZATION }}
\bigskip
\bigskip
\bigskip
\bigskip

\centerline{{\bf Eric D'Hoker}
\footnote{$^*$} {
 Research is supported in part by National Science Foundation grant
PHY-92-18990.}}

\medskip

\centerline{{ dhoker@physics.ucla.edu}}
\centerline{{\it Department of Physics and Astronomy}}
\centerline{{\it University of California at Los Angeles}}
\centerline{{\it Los Angeles, CA 90024, USA}}

\bigskip
\bigskip
\bigskip

\centerline{{\bf Abstract}}

\medskip

We propose actions for non-linear sigma models on cosets $G/H$ in 2+1
dimensions
that include the most general non-linear realizations of Chern-Simons terms.
When $G$ is simply connected and $H$ contains $r$ commuting U(1) factors, there
are $r$ different topologically conserved charges and generically
$r$ different types of topological solitons. Soliton spin fractionalizes as a
function of the Chern-Simons couplings, with independent spins associated to
each species of soliton charge, as well as to pairs of different
charges. This model of soliton spin fractionalization generalizes to arbitrary
$G/H$ a model of Wilczek and Zee for one type of soliton.

\bigskip
\bigskip

\vfill\eject

A complete characterization of invariant effective actions for Goldstone
boson fields due to the spontaneous breakdown of a symmetry group
$G$ to a subgroup $H$ was recently obtained in Ref. [1]. In addition to
manifestly $G$-invariant contributions arising from invariant Lagrangian
densities, there may in general be special terms in the action that do not
correspond to invariant Lagrangian densities.
These special actions were shown to
be in one to one correspondence with non-trivial generators of the de Rham
cohomology of $G/H$. They comprise the Wess-Zumino-Witten terms [2]
but may also receive contributions involving Goldstone-Wilczek currents [3].
For 2+1 dimensional space-time
$M_3$, the special classical actions are given by
$$
S[\pi] = \int _{B_4} \Omega (\tilde \pi)
\eqno (1)
$$
where $\Omega (\tilde \pi)$ is a cohomology generator of degree 4 belonging
to $H^4(G/H;{\bf R})$, $B_4$ is a four ball with boundary $M_3$ and $\tilde
\pi$
is a field on $B_4$ that continuously interpolates between the original
Goldstone field
$\pi(x) $ and the field 0.

In this letter, we determine explicitly the most general form of special
classical actions in 2+1 dimensions and the quantization conditions for the
associated coupling constants. Furthermore, when $G$ is simply connected and
$H$
contains
$r$ independent commuting $U(1)$ factors, there are $r$ different topologically
conserved currents with associated charges $Q_i$ and
$r$ different species of topological solitons. We derive
Bogomolnyi bounds on the masses and show that in general soliton spin
fractionalizes as a function of the non-linear Chern-Simons coupling constants
and the soliton charges. In addition to a fractional spin for each
species of soliton charge, there are also spin factors associated with pairs of
different charges. This model of soliton spin fractionalization generalizes  to
arbitrary $G/H$ the models considered for $G/H=S^2$ by Wilczek and Zee [4,5],
and for $G/H={\bf C}P^N$ in [7], where only one species of
solitons occurs. (See also [8] for applications in condensed matter physics.)

We begin by constructing the most general special actions of (1).
All cohomology generators were constructed in Ref. [9], and may be conveniently
expressed in terms of the components of the Maurer-Cartan connection defined by
$U^{-1}dU= \theta ^A T^A = \theta _\mu ^A T^A dx^\mu$. Here, $U( \pi)$
parametrizes the Goldstone field $\pi (x)$ and $T^A$ are the representation
matrices of $\G$ in the representation
\footnote {${}^\dagger$} {Indices $A$, $a$ and $\alpha$ run over the generators
of the Lie algebra $\G$ of $G$, the Lie algebra $\H$ of $H$ and the complement
$\M$ of $\H$ in $\G$ respectively, and $f_{\alpha \beta \gamma}$ denotes the
structure constants of $H$.}  of $U$. The most general cohomology generator of
degree 4 takes the form
$\Omega (\tilde \pi)=
\Omega _0 (\tilde \pi) + \Omega _1 (\tilde \pi) + \Omega _2 (\tilde \pi)$ with
$$
\eqalign{
\Omega _1(\tilde \pi) = &\sum _{\alpha, \beta =1} ^{{\rm dim H}}
m_{\alpha \beta}
W_\alpha W_\beta \qquad \qquad W_\alpha = d \theta _\alpha + \half f_{\alpha
\beta \gamma} \theta _\beta \theta _\gamma
\cr
\Omega _2(\tilde \pi) = & \sum _{\sigma _l} \mu ^{\{\sigma_l\}} _1
\theta ^{\sigma _1} \theta ^{\sigma_2}\theta ^{\sigma_3}\theta ^{\sigma_4}
+ \sum _{\sigma _1, \sigma _2, i}  \mu _2 ^{(\sigma _l, i)} \theta
^{\sigma _1} \theta ^{\sigma _2} W_{\alpha _i}
+ \sum _{\sigma, \rho }  \mu _3 ^{(\sigma, \rho)}\theta ^{\sigma}
\Omega _3^{(\rho)}
\cr}
\eqno (2)
$$
Here, $\Omega _0(\tilde \pi)$ is a generator with trivial cohomology which is
the exterior derivative of a manifestly invariant 3-form,
$W_\alpha$ is the field strength of the $\H$-valued connection
$V$ with components $V_\alpha =\theta _\alpha$ and $m_{\alpha \beta}$ is a
symmetric tensor on
$\H$, invariant under the adjoint action of $\H$. The differential
forms $\Omega _1 (\tilde \pi)$ generalize the well-known ~$\tr ~F\tilde F$ term
in four dimensions,
and may similarly be written as total exterior derivatives of
generalized  Chern-Simons terms [10,11]. The contribution
$\Omega _2 (\tilde \pi)$ arises when cohomology generators of degree 1, denoted
by $ \theta ^\sigma$,
$\sigma =1, \cdots , \dim H^1(G/H;{\bf R})$, are also present;
this  happens only
when $G/H$ is not simply connected. In the expression for
$\Omega _2(\tilde \pi)$, $\mu _1$, $\mu _2$ and $\mu _3$ are arbitrary
coefficients, anti-symmetric in their
$\sigma$ indices, and $\sigma$, $\alpha _i$ and $\rho$ running over the
generators of $H^1(G/H;{\bf R})$, $H^2(G/H;{\bf R})$ and $H^3(G/H;{\bf R})$
respectively.

The subgroup $H$ admits the following decomposition into $q$ connected, simply
connected, simple Lie groups $H_p$,
$p=1, \cdots, q$, times $r$ commuting $U(1)$ factors and a finite discrete
subgroup $D$ :
$$
H= H_1 \times \cdots \times H_q \times U(1) ^r \times D
\eqno (3)
$$
The invariant tensor $m_{\alpha \beta}$ decomposes under this product and
reduces to the Cartan-Killing form on the simple factors $H_p$.  As a result,
the effective actions in Eq. (1) may be rewritten directly in 2+1 dimensional
space-time, with $S[\pi]   =  S_0[\pi] + S_1[\pi] +S_2[\pi] $ and
$$
\eqalign{
S_1[\pi] = &   \sum _{p=1} ^q m_2 ^{p}\int _{M_3} \!\!
(V^{(p)} _\alpha dV^{(p)} _\alpha + {2 \over 3} f_{\alpha \beta \gamma} V^{(p)}
_\alpha V^{(p)}_\beta V^{(p)} _\gamma  )
+  \sum _{i,j=1} ^{r}  m_1 ^{ij} \int _{M_3} \!\! V_{\alpha _i} dV_{\alpha _j}
\cr
&\cr
S_2[\pi] = &  \sum _{\sigma _l} \mu ^{\{\sigma_l\}} _1 \int _{M_3} \!\!
\psi ^{\sigma _1} \theta ^{\sigma_2}\theta ^{\sigma_3}\theta ^{\sigma_4}
+ \sum _{i,\sigma _l}  \mu _2 ^{i~\sigma _l} \int _{M_3} \!\!
\psi ^{\sigma _1} \theta ^{\sigma _2} W_{\alpha _i}
+ \sum _{\sigma, \rho }  \mu _3 ^{\sigma \rho} \int _{M_3} \!\! \psi
^{\sigma}  \Omega _3^{(\rho)}
\cr}
\eqno (4)
$$
Here, $S_0[\pi]$ contains only manifestly invariant contributions, such as
kinetic terms with two derivative, Skyrme terms with four derivatives and
possibly higher order derivative terms. (We shall classify terms with two
derivatives shortly.) The contribution  $S_1[\pi]$ results from substituting
$\Omega [\tilde \pi]$ of Eq. (2) into the general expression for the action in
Eq. (1) and using the fact that $\Omega [\tilde \pi]$ is a total differential.
These terms correspond to Chern-Simons actions [10] evaluated on composite
gauge fields $V_\alpha$, and we shall name them non-linear Chern-Simons terms.
The contribution $S_2[\pi]$ is obtained from $\Omega _2(\tilde \pi)$ in (3)
with
the help of the angles $\psi ^\sigma$ that parametrize the homology cycles dual
to
$\theta ^\sigma = d\psi ^\sigma$.

The actions $S_1[\pi]$ and $S_2[\pi]$ are not in general gauge
invariant, even though their contributions to the Euler-Lagrange
equations are covariant. At the quantum level, the action is required to be
invariant only up to additive shifts which are integer multiples of $2
\pi i$. This condition in general requires certain of the coupling
constants $m_1$, $m_2$, $\mu _1$, $\mu _2$ or $\mu _3$
to be quantized in integer
multiples of some basic units [10,11]. Precisely which of these coupling
constants must be quantized depends upon the topology of space-time $M_3$.
If space-time is compactified to a sphere $S^3$, only the Chern-Simons
couplings
$m_2 ^{p}$ corresponding to the simple factors $H_p$ must be quantized,
while the couplings $m_1 ^{ij}$, $\mu _1$, $\mu _2$ and $\mu _3$ remain
unquantized. On the other hand, if space-time is  $M_3 = S^2 \times S^1$, the
couplings $m_1$ must be quantized, while the couplings $m_2$ remain
unquantized.
More generally, the couplings $m_1$ and $\mu _i$ must be quantized whenever
$M_3$ is not simply connected, while $m_2$ must be quantized if $\pi _3 (M_3)
\not=0$.

For simplicity, we shall henceforth assume $G$ simply connected,
$H$ connected (with the discrete subgroup $D=1$), so that $G/H$ is simply
connected,
$H^1(G/H;{\bf R})=0$ and the contribution $S_2[\pi]$
vanishes identically. As a result, $H^2(G/H;{\bf Z}) \sim \pi _2(G/H)= \pi
_1(H)=\oplus _r {\bf Z}$,
and there exist $r$ independent topologically conserved
currents which are proportional to the $U(1)^r$ field strengths $W_{\alpha
_l}=dV_{\alpha _l}$. Their associated conserved topological charges are given
by
$$
Q_i = \int _{M_2} W_{\alpha _i} \qquad \qquad i=1, \cdots , r
\eqno (5)
$$
Gauging these topological currents yields an equivalent representation of the
special action $S_1$ of (4), considered here just for $m_2=0$. To see
this, we consider
$$
\tilde S_1 [\pi, B] = 2\sum _{i=1} ^r  \int _{M_3} B^i W_{\alpha _i}
-\sum _{i,j=1} ^r g_{ij} \int _{M_3} B^i dB^j
\eqno (6)
$$
The associated Euler-Lagrange  equation for $B^i$ can be solved explicitly in
terms of $V_{\alpha _j}$ and arbitrary gauge transformation functions $\omega
^i$ :
$$
B^i_0 = \sum _{j=1} ^r (g^{-1})^{ij} V_{\alpha _j} + d \omega ^i
\eqno (7)
$$
Thus, $B^i$ is not a dynamical field, even though it enters (6) with
derivatives. It may be eliminated from $\tilde S_1$, yielding precisely
$S_1[\pi]$ (again for $m_2=0$) provided
$g_{ij}$ is the inverse matrix of $m_1^{ij}$. $\tilde S_1$ may be viewed as an
ordinary (linear) Chern-Simons interaction for gauge fields coupled to
topological currents.

Topological solitons, labelled by $r$ independent charges $Q_i$, $i=1,\cdots,r$
arise in these models. Mathematically, these solitons are identical to 1+1
dimensional instantons on coset spaces $G/H$ [12].
Physically, they strongly resemble `t~Hooft-Polyakov magnetic
monopoles in 3+1 dimensions [13], produced by the spontaneous breaking of a
semi-simple group $G$ to a subgroup $H$ containing $r$ commuting
(electro-magnetic) $U(1)$ factors. (See also [14].)
The precise shape of the soliton solution depends upon the nature of the
action $S_0[\pi]$ and on the group theory of
$G/H$. Here, we shall consider only the two-derivative kinetic contributions to
$S_0[\pi]$, but higher order terms are easily included. As the properties
of solitons depend primarily on the $U(1)$ factors in $H$, we omit
discussion of the $m_2$ term in $S_1$.

We evaluate the following soliton quantum numbers : topological
charges, mass and spin. We denote the Cartan generators of
$\G$ by $h_i$, $i=1,\cdots, {\rm rank}~G$, of which $i=1,\cdots, r$
generate $U(1)^r$ in $\H$.  Roots $a$ are associated with generators $T_a$
with $[h_i, T_a ] = a_i T_a $ and $[T_a, T_{-a}]=\sum _i a_i h_i$.
The conserved topological charges can be expressed as a sum over positive roots
$$
Q_i = -i \sum _{a>0} a_i \int _{M_2} \!\! d^2x ~ \epsilon ^{mn} \theta ^a _m
\theta _n ^{-a} \qquad \qquad i=1,\cdots, r
\eqno (8)
$$
They take on integer values provided $\theta _m ^a$ falls off sufficiently
rapidly as $|\vec x|\to \infty$. The most general kinetic action with just two
derivatives can be expressed as
$$
S_0[\pi] = \sum _{a,b} {1\over 2} K_{ab} \int _{M_3} \!\! d^3x~ \theta _\mu ^a
\theta _\mu ^b
\eqno (9)
$$
Here, $K_{ab}$ is symmetric with positive eigenvalues, and $U(1)^r$ invariance
requires $a_i +b _i =0$ for all $i=1,\cdots, r$. When $r<{\rm rank} ~G$, the
sum over $a$ and $b$ may in principle also contain terms from Cartan
generators $h_i$, $i=r+1, \cdots, {\rm rank} ~G$ which are not in $U(1)^r$. In
particular, when non-trivial simple components $H_p$ are present in
$H$ in addition to $U(1)^r$, $K_{ab}$ should be invariant under $H_p$.
Henceforth, to simplify the discussion, we shall assume that $K_{ab}$ vanishes
unless $a$ and $b$ are opposite roots (This is always the case when rank $G$
= rank ~$H$.) and $K_{a~-a}\geq 0$ for all positive roots $a$.

The masses of topological solitons obey Bogomolnyi type bounds [15] at the
classical level. To derive those,  we start from the following identity,
valid for a fixed root $T_a$. ($m,n=1,2$ run over space coordinates only.)
$$
\int _{M_2} \!\!\! d^2x~ \theta _m ^a \theta _m ^{-a} ={1 \over 2} \int _{M_2}
\!\!\! d^2x ~(\theta _m ^a \pm i \epsilon _{mn} \theta _n ^a)
(\theta _m ^{-a} \mp i \epsilon _{mp} \theta _p ^{-a})
\pm i \int _{M_2} \!\!\! d^2 x~ \epsilon _{mn} \theta _m ^a \theta _n ^{-a}
\eqno (10)
$$
Let $\vec \lambda$ be a vector with components $\lambda _i$ which is
independent of the roots $a$ and such that for all positive roots,
$$
K_{a~-a} = \vec \lambda \cdot \vec a + N_a
\qquad \qquad
N_a \geq 0
\qquad\qquad
\vec \lambda \cdot \vec a \geq 0
\eqno (11)
$$
(The vector $\vec \lambda$ is naturally
proportional to a vector in the lattice dual to the roots $\vec a$,
which can be viewed as a highest weight vector of some
irreducible representation of $G$.)
For a given set of constants $K_{a~-a}$, there may be several possible choices
for $\vec  \lambda$. For each choice, we obtain the following bound on the
energy $E[\pi]$ of any static configuration $\pi (x)$ :
$$
\eqalign{
 E[\pi] = & \sum _{a>0} N_a \int _{M_2} \!\!\!d^2x ~\theta _m^a \theta _m ^{-a}
+{1 \over 2} \sum _{a >0} \vec \lambda \cdot \vec a \int _{M_2} \!\!\! d^2x
{}~\big |\theta _m ^a \pm i \epsilon _{mn} \theta _n ^a \big |^2 \mp \vec
\lambda
\cdot \vec Q
\cr
\geq & |\vec \lambda \cdot \vec Q |
\cr}
\eqno (12)
$$
The lower bound can be saturated under the following conditions on the various
roots : if $\mp \vec \lambda \cdot \vec Q >0$, then
$$
\cases{
N_a = & $0 \qquad \qquad
\theta _m^a \pm i \epsilon _{mn} \theta _n ^a =0$ \cr
&\cr
N_b\not=& $0 \qquad \qquad
\theta _m ^b=0$ \cr}
\eqno (13)
$$
Thus, the Maurer-Cartan form $\theta _m$ lives in a subalgebra of $\G$ in
which all roots have $N_a=0$.

Solutions may be found explicitly in terms of analytic functions using
techniques of [12,14]. The simplest case is when the soliton lives in an
$SU(2)$ subgroup of $G$ which contains a generator of the Abelian $U(1)$ factor
$U(1)^r$ in $H$. Then, the $SU(2)$ subalgebra is spanned by a Cartan generator
$h\in U(1)^r$ and by the roots $T_c$ and $T_{-c}$. The Maurer-Cartan form for a
configuration that satisfies the Bogomolnyi bound vanishes for all roots except
$c$ and $-c$, and we have
$$
\theta _m= \theta _m ^h h + \theta _m ^c T_c + \theta _m ^{-c} T_{-c}
\eqno (14)
$$
The Cauchy-Riemann equations  of (13) reduce to $\theta _{\bar z} ^c =0$, or
equivalently $\theta _z ^{-c} =0$. This equation is easily solved as follows :
$$
U(z,\bar z) = \exp \{ f^c (z) T_c\} \cdot \exp \{ f^{-c} (z, \bar z) T_{-c} \}
\cdot \exp \{ f^h (z,\bar z) h \}
\eqno (15)
$$
Here $f^{-c}$ and $f^h$ are arbitrary functions of $z$ and $\bar z$, $f^c$ is
holomorphic in $z$, and one imposes the additional requirement that
$U(z,\bar z)$ be in $\G$. For $SU(2)$ in the fundamental representation for
example, we have
$$
U(z, \bar z) = \pmatrix{1&0\cr if(z) & 1\cr}
\cdot \pmatrix{\rho (z,\bar z) & i \overline{f(z)} \rho (z,\bar z) \cr
0 & \rho ^{-1}(z,\bar z) \cr}
\cdot \pmatrix{ e^{i\varphi (z,\bar z) } & 0 \cr
0 & e^{-i\varphi (z,\bar z)} \cr}
\eqno (16)
$$
with the requirement that $(1 +|f(z)|^2 ) \rho ^2 (z,\bar z)=1$. The
corresponding soliton solutions satisfy $Q_{i_0}=1$ and $Q_i=0$ for
$i\not=i_0$, and are {\it elementary solitons} with the lowest possible
non-trivial charge assignment.

The position of the Bogomolnyi soliton inside $G/H$ may however be more general
than considered above. It is determined by the position of the homology 2 cycle
in $G/H$ dual to the closed invariant 2 forms $W_{\alpha_i}$. For example, in
the case of $G/H=SU(3)/U(1)$ we take $H=U(1)$ the subgroup generated by the Lie
algebra generator $\lambda _8 \sim {\rm diag}(1~~1~-2)$, which does not belong
to any $SU(2)$ subalgebra of $SU(3)$. There is only a single conserved current,
$W_{\lambda _8}$, which uniquely determines the soliton charge. Clearly,
$W_{\lambda _8}$ is invariant under local $SU(2)$ right  multiplications of $U$
that commute with $\lambda _8$. As a result, $W_{\lambda _8}$ projects down to
a
well-defined closed 2-form on the coset of $G/H$ by
$SU(2)$ which is ${\bf C}P^2 = SU(3) /S(U(2)\times U(1))$. On this space,
$W_{\lambda _8}$ is just the K\"ahler class, whose structure is well-known
[16].
More generally, when $r<{\rm rank}~G$, one may use the methods that generalize
our discussion for $SU(3)$, by forming cosets of $G/H$ by subgroups of $G$ that
leave the relevant $W_{\alpha _i}$ invariant, until one arrives at a space
$G/H'$ with ${\rm rank }~H'={\rm rank}~G$. These spaces can be viewed as
classifying spaces [17] for the second cohomology classes of $G/H$.

The Bogomolnyi bounds guarantee that the solitons obey the field equations, but
they allow for the possibility that some solutions correspond to multi-particle
configurations. In general, the analysis of stability is complicated by the
arbitrariness of the kinetic energy coefficients $K_{a~-a}$ in (9), upon whose
values the stability questions depend. In the simplest case,  ${\rm
rank}~G=r$, elementary solitons with $Q_{i_0}=1$ and $Q_i=0$ for $i\not=i_0$,
are all stable. Interaction energies between different elementary solitons
vanish when all $N_a=0$. Thus, all
non-elementary configurations with non-negative charges $Q_i$ are
multi-solitons
composed of non-interacting elementary solitons. This property is very much
analogous to that of multi-monopole configurations. Elementary solitons and
anti-solitons (with negative elementary charge $Q_i$) presumably attract one
another, but this conclusion does not follow from the Bogomolnyi bounds.

The soliton spin was evaluated in [4,5] for $G/H=S^2$ by considering the change
in the action under a rotation of the soliton configuration. Generalization
of this method to arbitrary $G/H$ is straightforward only when the solitons are
embedded into $G$ in a simple way, as discussed above. Parametrizations of
solitons in more general embeddings however is cumbersome.
Fortunately, a simple determination of soliton spin is available
directly from current algebra [6] at the classical level,
which we shall make use
of here. We set $S_2[\pi]=0$ and $m_2=0$, so that the only remaining
non-linear Chern-Simons action is the $m_1$ term in $S_1[\pi]$.
All other contributions to the
effective action contained in $S_0[\pi]$ are manifestly invariant. To fix
ideas, we may think of this action as just the kinetic term, given in (9).
Following [6], the definition of spin is through the stress tensor
$$
J=\epsilon _{mn} \int _{M_2}\!\!\! d^2x~ x^m T^{0n}
\eqno (17)
$$
The stress tensor $T^{\mu \nu}$ is constructed as the variation of the action
under the cahnge of an auxiliary
background space-time metric tensor, and thus receives no contribution from
the topological part of the action, $S_1[\pi]$. We find
$$
T^{0n} = \partial _n \pi ^a \Pi ^a _0
\qquad\qquad
\Pi ^a _0 = {\partial {\cal L}_0 \over \partial (\partial _0 \pi ^a)}
\eqno (18)
$$
where $\Pi^a_0 (x) $ is the part of the momentum canonically conjugate to
$\pi ^a$, arising only from the action $S_0$ with
Lagrangian density ${\cal L}_0$. It is related to the full canonical momentum
$\Pi^a$, by the addition of a term involving only the action $S_1$ :
$$
\Pi^d (x) = \Pi ^d _0 (x) + \sum _{i,j} m_1 ^{ij} \epsilon _{mn} f^{\alpha _i
bc} M^{bd} (\pi (x)) \theta _m ^c (x) V_n ^{\alpha _j}(x)
\eqno (19)
$$
where $M^{bd}$ is defined by $\theta ^b _m = M^{bd} \partial _m \pi ^d$.
One proceeds by expressing the composite gauge field $V_n ^{\alpha _j} $ in
terms of the topologically conserved current $W^{\alpha _j}$. It is then
straightforward to work out the formula for the contribution to the spin
arising from the non-linear Chern-Simons term $S_1$, and we find
$$
J = \int _{M_2} \!\!\! d^2x~ \epsilon _{mn} x_m \Pi ^a(x) \partial _n \pi ^a(x)
- {1 \over 2 \pi} \sum _{i,j} m_1 ^{ij} Q_i Q_j
\eqno (20)
$$
The first term on the right hand side is the standard
spin, which in canonical quantization receives only
integer contributions. The second term arises purely from the
non-linear Chern-Simons term. While the values of $Q_i$ are integers for
configurations that fall off sufficiently fast at spatial infinity, the
coupling constants $m_1 ^{ij}$ do not have to be quantized in general. As a
result, the non-linear Chern-Simons contribution to the spin $J$ is in general
fractional. When there is only a single topologically conserved charge, we
recover the expression of [6]. However, for several different conserved
charges,
the off diagonal entries in $m_1 ^{ij}$ contribute and  add spin
from the presence of several different charges $Q_i$.

\bigskip
\bigskip

\noindent
{\bf Acknowledgements}

I have benefitted from helpful conversations with Roman Jackiw, Jeff Rabin,
Terry Tomboulis, Steven Weinberg and Tony Zee.

\bigskip
\bigskip

\noindent
{\bf References}

\medskip

\item{[1]} E. D'Hoker and S. Weinberg, Phys. Rev. {\bf D50} (1994) 605.

\item{[2]}J. Wess and B. Zumino, Phys. Lett. {\bf 37B} (1971) 95; E. Witten,
Nucl. Phys. {\bf B223} (1983) 422, 433.

\item{[3]} J. Goldstone and F. Wilczek, Phys. Rev. Lett. {\bf 47}
(1981) 986; E. D'Hoker and J. Goldstone, Phys. Lett. {\bf 158B} (1985) 429.

\item{[4]}F. Wilczek and A. Zee, Phys. Rev. Lett. {\bf 51} (1983) 2250;
Y.S. Wu and A. Zee, Phys. Lett. {\bf 147B} (1984) 325.

\item{[5]} F. Wilczek, {\it Fractional Statistics and Anyon
Superconductivity}, World Scientific, Singapore, 1990; E. Fradkin, {\it Field
Theories of Condensed Matter Systems}, Addison Wesley, New York, 1991; W.J.
Zakrzewski, {\it Low Dimensional Sigma Models}, Adam Hilger, London 1989.

\item{[6]} M.J. Bowick, D. Karabali, L.C.R. Wijewardhana, Nucl. Phys. {\bf
B271}
(1986) 417.

\item{[7]}N.K. Pak, Phys. Lett. {\bf 260B} (1991) 377;
G. Ferretti and S.G. Rajeev, Phys. Rev. Lett. {\bf 69} (1992) 2033; G.
Ferretti, S.G. Rajeev and Z. Yang, Int. J. Mod. Phys. {\bf A7} (1992) 7989; K.
Fujii, Commun. Math. Phys. {\bf 162} (1994) 273; K. Fujii, J. Math. Phys. {\bf
36} 97; S. Forte, Rev. Mod. Phys. {\bf 64} (1992) 193; A.P. Balachandran, M.
Bourdeau and S. Jo, Int. J. Mod. Phys. {\bf A5} (1990) 2423.

\item{[8]}Y.-S. Wu, Phys. Rev. Lett. {\bf 52} (1984) 2103; D.P.~Arovas,
J.R.~Schrieffer, F.~Wilczek and A.~Zee, Nucl. Phys. {\bf B251} (1985) 117;
D.-H.~Lee and C.L.~Kane, Phys. Rev. Lett. {\bf 64} (1990) 1313; S.L.~Sondhi,
A.~Karlhede, S.A.~Kivelson and E.H.~Rezayi, Phys. Rev. {\bf B47} (1993) 16419;
K.~Moon, H.~Mori, K.~Yang, S.M.~Girvin, A.H.~MacDonald, L.~Zheng and
D.~Yoshioka, Phys. Rev. {\bf 51} (1995) 5138; C.~Nayak and F.~Wilczek,
PUPT~1540, IASSNS-HEP~95/35 preprint, May 1995.

\item{[9]} E. D'Hoker, ``Invariant Effective Actions, Cohomology of Homogeneous
Spaces and Anomalies", UCLA/95/TEP/5 preprint (1995), hep-th/9502162, accepted
for publication in Nucl. Phys. B.; E. D'Hoker, ``Invariant Effective Actions
and
Cohomology", UCLA/95/TEP/11 preprint (1995), hep-th/9505109.

\item{[10]} S. Deser, R. Jackiw and S. Templeton, Phys. Rev. Lett. {\bf 48}
(1982) 975; Ann. Phys. (NY) {\bf 140} (1982) 372; J. Schonfeld, Nucl. Phys.
{\bf B185} (1981) 157; W. Siegel, Nucl. Phys. B (1982).

\item{[11]}R. Jackiw, in {\it Current Algebra and
Anomalies}, by S.B. Treiman, R. Jackiw, B. Zumino and E. Witten, Princeton
Univ.
Press, 1989.

\item{[12]}A. Belavin and A.M. Polyakov, JETP Lett. {\bf 22} (1975) 245; V.
Golo and A. Perelemov, Phys. Lett. {\bf 79B} (1978) 112; A. D'Adda, P. Di
Vecchia and M. L\"uscher, Nucl. Phys. {\bf B146} (1978) 63; A. Perelemov, Phys.
Rep. {\bf 146} (1987) 135.

\item{[13]} G. `t Hooft, Nucl. Phys. {\bf B79} (1974) 276; A.M. Polyakov, JETP
Lett. {\bf 20} (1974) 194.

\item{[14]} R. Jackiw, K. Lee, E. Weinberg, Phys. Rev. {\bf D42} (1990) 3488;
R. Jackiw and S.-Y. Pi, Phys. Rev. Lett. {\bf 64} (1990) 2969.

\item{[15]} E. Bogomolnyi, Sov. J. Nucl. Phys. {\bf 24} (1976) 449.

\item{[16]} B.A. Dubrovin, A.T.~Fomenko and S.P.~Novikov, {\it Modern
Geometry and Applications}, Springer Verlag, 1990.

\item{[17]} N. Steenrod, {\it The Topology of Fibre Bundles}, Princeton U.
Press, 1951.

 \end